\documentstyle[amsmath,epsf,a4,11pt]{article}
\begin{document}

\date{}

\title{{\bf Structure and stability of the Lukash plane-wave
spacetime}}\author{John D. Barrow\thanks{e-mail address:
j.d.barrow@damtp.cam.ac.uk} and Christos G. Tsagas\thanks{e-mail
address: c.tsagas@damtp.cam.ac.uk}\\ {\small DAMTP, Centre for
Mathematical Sciences, University of Cambridge, Wilberforce
Road}\\ {\small Cambridge~CB3~0WA, UK}}

\maketitle

\begin{abstract}
We study the vacuum, plane-wave Bianchi $VII{}_{h}$ spacetimes
described by the Lukash metric. Combining covariant with
orthonormal frame techniques, we describe these models in terms of
their irreducible kinematical and geometrical quantities. This
covariant description is used to study analytically the response
of the Lukash spacetime to linear perturbations. We find that the
stability of the vacuum solution depends crucially on the
background shear anisotropy. The stronger the deviation from the
Hubble expansion, the more likely the overall linear instability
of the model. Our analysis addresses rotational, shear and Weyl
curvature perturbations and identifies conditions sufficient for
the linear growth of these distortions.\\\\ PACS Numbers:
98.80.Jk, 98.80.Cq, 98.80.Bp
\end{abstract}

\section{Introduction}
%%%%%%%%%%%%%%%%%%%%%%
There has long been an interest in the study of the spatially
homogeneous Bianchi spacetimes and their cosmological applications
to our understanding of singularities and of the observed level of
isotropy in the universe. These studies analyse the problems
within the manageable domain of ordinary differential equations
and provide only a finite number of alternative cosmologies
(see~\cite{HW1} and references therein). The most general Bianchi
universes that contain the open Friedmann model as a special
subcase are those of type $VII{}_{h}$. The late-time asymptotes
for the non-tilted type $VII{}_{h}$ spacetimes, with $h\neq0$ and
a matter content that obeys the strong energy condition, evolve
towards the vacuum plane-wave solution found by Doroshkevich et al
and Lukash~\cite{DLN,L2} that is known as the Lukash metric. These
spacetimes describe the most general effects of spatially
homogeneous perturbations on open Friedmann universes; see for
example~\cite{CH}-\cite{BT}. The Lukash metric plays a guiding
role in these investigations because of the subtle stability
properties of isotropic expansion at late times in open universes.
When the strong energy condition is obeyed, then isotropic
expansion was found to be stable but not asymptotically stable at
late times~\cite{B1}-\cite{BT}.

Traditionally, Bianchi spacetimes have been studied qualitatively,
primarily by means of dynamical system methods~\cite{C}-\cite{WL}.
The same techniques also facilitate the analysis of the less well
understood tilted Bianchi models, namely those where the fluid
4-velocity is no longer orthogonal to the hypersurfaces of
constant time~\cite{KE}-\cite{HvdHC}. In this paper we attempt an
analytical approach. In particular, we combine covariant and
orthonormal frame techniques to provide a description of the
Lukash spacetime of Bianchi type $VII{}_{h}$ in terms of its
irreducible kinematical and geometrical quantities. We then use
the zero-order results to study inhomogeneous perturbations around
the vacuum solution and discuss its linear stability. In so doing,
we allow for the presence of a low-density matter component with a
pressure-free equation of state. Our main interest is the
evolution of perturbations in the kinematics and the geometry of
the vacuum model and its linear response to these distortions. We
find that stability depends primarily on the amount of the
background shear anisotropy. In particular, our results show that
the higher the background shear the more likely is the linear
instability of the model. Another key factor is the relative
orientations of the various kinematical and geometrical
quantities. The positioning of the vorticity vector with respect
to the principal axes of shear, for example, or the relative
orientation between the shear and the spatial curvature
eigenframes can also influence the linear stability of the Lukash
spacetime.

We also consider rotational perturbations and study their
evolution relative to the background volume expansion. Our
analysis provides a condition for the linear growth of vortical
distortions that depends primarily on the background shear
anisotropy. We identify, in particular, the minimum amount of
shear necessary for the linear instability of the Lukash universe
against vorticity perturbations. The level of background shear can
also determine whether kinematical anisotropies will remain
bounded or dominate the linear expansion of the perturbed Lukash
model. Additionally, by allowing for low-density dust matter, we
find that its presence has no effect on the vorticity and shear of
the perturbed spacetime. In contrast, the introduction of a
material component can have an effect on the linear Weyl
anisotropy of the model. More specifically, the presence of even a
low-density dust fluid seems enough to ensure that the Lukash
solution will diverge from its original plane-wave nature at late
times.

\section{The Lukash plane-wave attractor}
%%%%%%%%%%%%%%%%%%%%%%%%%%%%%%%%%%%%%%%%%
The Bianchi $VII{}_{h}$ models belong to the non-exceptional
family of the Behr class B spatially homogeneous spacetimes. The
plane-wave Lukash solution is the late-time attractor of the
Bianchi $VII{}_{h}$ models for a broad range of initial date and
matter properties. These vacuum spacetimes correspond to
equilibrium points of the associated autonomous dynamical system
and are self-similar~\cite{Ea}-\cite{A2}. The line element of the
Lukash metric takes the form
\begin{equation}
{\rm d}s^{2}=-{\rm d}t^{2}+ t^{2}{\rm d}x^{2}+ t^{2r}{\rm e}^{2rx}
\left[(A{\rm d}y+B{\rm d}z)^{2}+(C{\rm d}y+A{\rm
d}z)^{2}\right]\,,  \label{DLN}
\end{equation}
where $r$ is an arbitrary constant parameter in the range $0<r<1$,
$A=\cos v$, $B=f^{-1}\sin v$, $C=-f\sin v$ and $v=k(x+\ln
t)$~\cite{HW,HSUW}. Note that $f$ and $k$ are constants related to
$r$ by
\begin{equation}
\frac{k^{2}}{f^{2}}(1-f^{2})^{2}=4r(1-r) \hspace{15mm} {\rm and}
\hspace{15mm} r^{2}=hk^{2}\,,  \label{Lcons}
\end{equation}
where $h$ is the associated group parameter. As we shall see next,
constraint (\ref{Lcons}) is the Lukash analogue of the Friedmann
equation. We also point out that for $r=1$ and $f^{2}=1$ the
Lukash metric reduces to that of the empty Milne universe.

\section{Covariant description}
%%%%%%%%%%%%%%%%%%%%%%%%%%%%%%%
Consider a family of observers, with worldlines tangent to the
timelike velocity field $u_{a}$ (normalised so that
$u_{a}u^{a}=-1$). The latter, together with the associated
projection tensor $h_{ab}=g_{ab}+u_{a}u_{b}$, introduces a local
1+3 threading of the spacetime into time and space. One can then
decompose the various kinematical, dynamical and geometrical
quantities into their respective irreducible parts and obtain a
completely covariant description of the spacetime~\cite{E1,EvE}.

\subsection{Covariant variables}
%%%%%%%%%%%%%%%%%%%%%%%%%%%%%%%%
The covariant formalism uses the irreducible kinematic quantities,
the energy density and pressure of the matter fields and the
gravito-electromagnetic tensors, instead of the metric which in
itself does not provide a covariant description. The key equations
are the Ricci and Bianchi identities, applied to the observers'
4-velocity, while Einstein's equations are incorporated via
algebraic relations between the Ricci and the matter
energy-momentum tensors. Thus, in the absence of matter and
rotation, the plane-wave attractors of the Bianchi $VII{}_{h}$
spacetimes are covariantly characterised by\footnote{Throughout
this article we employ a Lorentzian metric with signature
($-,\,+,\,+,\,+$) and use geometrised units with $c=1=8\pi G$.
Consequently, all geometrical variables have physical dimensions
that are integer powers of length. Also, Latin indices take the
values 0,1,2,3 and Greek ones run from 1 to 3.}
\begin{equation}
\mu=0=p=q_{a}=\pi_{ab} \hspace{15mm} {\rm and} \hspace{15mm}
\dot{u}_{a}=0=\omega_{a}\,,  \label{Lccon1}
\end{equation}
while
\begin{equation}
\Theta,\,\sigma_{ab},\,E_{ab},\,H_{ab}\neq 0\,.  \label{Lccon2}
\end{equation}
Note that $\mu,\,p,\,q_{a}$ and $\pi_{ab}$ are respectively the
energy density, the isotropic pressure, the heat flux and the
anisotropic stresses of the matter,
$\Theta,\,\sigma_{ab},\,\omega_{a}$ and $\dot{u}_{a}$ are the
volume expansion, the shear, the vorticity and the acceleration,
while $E_{ab}$ and $H_{ab}$ are the electric and magnetic parts of
the Weyl tensor $(C_{abcd})$. The latter have equal magnitudes and
are orthogonal to each other, in accord with the Petrov type N
nature of the Lukash solution. In other words,
\begin{equation}
E^{2}=H^{2} \hspace{15mm} {\rm and} \hspace{15mm}
E_{ab}H^{ab}=0\,, \label{Lccon3}
\end{equation}
where $E^{2}=E_{ab}E^{ab}/2$ and $H^{2}=H_{ab}H^{ab}/2$. The
former of these constraints implies that $C_{abcd}C^{abcd}=0$. The
latter ensures that $C_{abcd}{}^{\ast }C^{abcd}=0$, where
${}^{\ast}C_{abcd}=\eta_{abqs}C^{qs}{}{}_{cd}/2$ and $\eta
_{abcd}$ is the 4-dimensional alternating tensor. Note that the
Weyl curvature invariant $C_{abcd}C^{abcd}$ has been suggested and
used as a measure of the gravitational entropy by several
authors~\cite{P}-\cite{BH2}, but cannot on its own capture
deviations from isotropic expansion in plane-wave spacetimes.

\subsection{Covariant equations}
%%%%%%%%%%%%%%%%%%%%%%%%%%%%%%%%
The average volume expansion of the Lukash universe is described
by the following version of the Raychaudhuri equation
\begin{equation}
\dot{\Theta}=-{\textstyle{\frac{1}{3}}}\Theta^{2}-2\sigma^{2}\,,
\label{Ray1}
\end{equation}
where $\sigma^{2}=\sigma_{ab}\sigma^{ab}/2$ is the magnitude of
the shear tensor. As usual, the expansion scalar is used to define
an average scale factor ($a$) via the standard relation
$\Theta/3=\dot{a}/a$.

The absence of matter means that the Lukash spacetime is Ricci
flat. The curvature of the spatial sections, however, is not zero.
In particular, zero rotation ensures that the 3-Ricci tensor
(${\cal R}_{ab}$) is completely determined by its scalar and its
symmetric and trace-free parts. These are given respectively by
\begin{eqnarray}
{\cal R}&=& -{\textstyle{\frac{2}{3}}}\Theta^{2}+ 2\sigma^{2}\,,
\label{LcR}\\ {\cal S}_{ab}&=&
-{\textstyle{\frac{1}{3}}}\Theta\sigma_{ab}+ \sigma_{c\langle
a}\sigma^{c}{}_{b\rangle}+ E_{ab}\,, \label{LcSab}
\end{eqnarray}
where ${\cal S}_{ab}={\cal R}_{\langle ab\rangle}={\cal
R}_{(ab)}-{\cal R}h_{ab}/3$.\footnote{Angled brackets denote the
symmetric and trace-free part of orthogonally projected tensors
and the orthogonally projected components of vectors.} As we will
see below, the scalar ${\cal R}$ is negative, which means that the
model is spatially open. Note that expression (\ref{LcR}) is the
generalised Friedmann equation.

In covariant terms, gravitational waves are described by the
electric and the magnetic parts of the Weyl tensor. The latter
obey a set of three coupled propagation equations, which are
accompanied by an equal number of constraints. In the case of the
Lukash plane-wave spacetime the evolution equations take the form
\begin{eqnarray}
\dot{E}_{ab}&=& -\Theta E_{ab}+ {\rm curl}H_{ab}+
3\sigma_{c\langle a}E^{c}{}_{b\rangle}\,,  \label{dotE}\\
\dot{H}_{ab}&=& -\Theta H_{ab}- {\rm curl}E_{ab}+
3\sigma_{c\langle a}H^{c}{}_{b\rangle}\,,  \label{dotH}\\
\dot{\sigma}_{ab}&=& -{\textstyle{\frac{2}{3}}}\Theta\sigma_{ab}-
E_{ab}- \sigma_{c\langle a}\sigma^{c}{}_{b\rangle}\,.
\label{dotsigma}
\end{eqnarray}
The constraints, on the other hand, are
\begin{eqnarray}
{\rm D}^{b}\sigma_{ab}&=& {\textstyle{\frac{2}{3}}}{\rm
D}_{a}\Theta\,,  \label{con1}\\ {\rm D}^{b}E_{ab}&=&
\epsilon_{abc}\sigma^{b}{}_{d}H^{cd}\,,  \label{con2}\\ {\rm
D}^{b}H_{ab}&=& -\epsilon_{abc}\sigma^{b}{}_{d}E^{cd}\,,
\label{con3}
\end{eqnarray}
where $\epsilon_{abc}=\eta_{abcd}u^{d}$ is the spatial alternating
tensor. In addition, the shear and the magnetic component of the
Weyl tensor are directly related by
\begin{equation}
H_{ab}={\rm curl}\sigma_{ab}\,,  \label{con4}
\end{equation}
with ${\rm curl}H_{ab}=\epsilon_{cd\langle a}{\rm
D}^{c}H^{d}{}_{b\rangle}$ by definition. Clearly, exactly
analogous expressions define ${\rm curl}E_{ab}$ and ${\rm
curl}\sigma_{ab}$. Note that the presence of (standing)
gravitational waves is guaranteed by the non-zero values of both
${\rm curl}E_{ab}$ and ${\rm curl}H_{ab}$. This is possible,
despite the spatial homogeneity of the Lukash spacetime, because
of the non-zero 3-curvature of the model. In other words, ${\rm
D}_{c}E_{ab}\,,\;{\rm D}_{c}H_{ab}\neq0$ due to non-zero
Christoffel symbols.

\section{Orthonormal-frame description}
%%%%%%%%%%%%%%%%%%%%%%%%%%%%%%%%%%%%%%%
The orthonormal frame formalism is an 1+3 decomposition of the EFE
into evolution and constraint equations relative to the timelike
vector field ${\bf e}_0$ of an orthonormal frame $\{{\bf
e}_a\}$~\cite{EMc}-\cite{ESW}. In cosmological studies ${\bf e}_0$
is the fundamental 4-velocity field, usually identified with the
motion of the cosmic medium. In models that accept an isometry
group, ${\bf e}_0$ can also be chosen as the normal to the
spacelike group orbits.

\subsection{Structure constants}
%%%%%%%%%%%%%%%%%%%%%%%%%%%%%%%%
The Lukash solution belongs to the non-exceptional family of the
Bianchi class B spacetimes. For these models the structure
constants $n_{\alpha\beta}$ and $a_{\alpha}$ (with
$n_{\alpha\beta}a^{\alpha}=0$) take the form\footnote{In the
orthonormal formalism the spacetime metric is
$g_{ab}=\eta_{ab}={\rm diag}\,(-1,1,1,1)$ and the spatial frame
vectors are $\{{\bf e}_{\alpha}\}$. Greek indices are raised and
lowered by means of the spatial metric
$g_{\alpha\beta}=\delta_{\alpha \beta }$.}
\begin{equation}
n_{\alpha\beta}= {\rm diag}\,(0,\,n_{2},\,n_{3}) \hspace{15mm}
{\rm and} \hspace{15mm} a_{\alpha}= (a_{1},\,0,\,0)\,.  \label{SC}
\end{equation}
Moreover, the self-similarity of the Bianchi $VII{}_{h}$
plane-wave attractor guarantees that the three non-zero components
of the structure constants are given by~\cite{HW}
\begin{equation}
a_{1}=-\frac{r}{t}\,, \hspace{5mm} n_{2}=\frac{k}{ft}
\hspace{15mm} {\rm and} \hspace{15mm} n_{3}=\frac{kf}{t}\,,
\label{a1n2n3}
\end{equation}
where $r$, $k$ and $f$ are constants related with each other and
with the group parameter of the model (see Eq.~(\ref{Lcons})).

\subsection{Kinematics}
%%%%%%%%%%%%%%%%%%%%%%%
Non-exceptional, non-tilted Bianchi class B spacetimes, like the
Bianchi $VII{}_{h}$ cosmologies, have $\sigma_{12}=0=\sigma_{13}$.
Hence, the self similarity of the vacuum plane-wave attractor of
these models means that the remaining components of the shear
tensor are
\begin{equation}
\sigma_{11}=\frac{2(1-r)}{3t}\,, \hspace{5mm}
\sigma_{22}=\sigma_{33}=-\frac{1-r}{3t} \hspace{15mm} {\rm and}
\hspace{15mm} \sigma_{23}=\frac{k(1-f^{2})}{2ft}\,.
\label{Lsigma}
\end{equation}
On using the above, one finds that the magnitude of the shear
tensor associated with the Lukash solution is
\begin{equation}
\sigma^{2}= {\textstyle{\frac{1}{2}}}\sigma_{\alpha\beta}
\sigma^{\alpha\beta}= \frac{(1-r)(1+2r)}{3t^{2}}\,,
\label{Lsigma2}
\end{equation}
where $\sigma_{\alpha\beta}$ are the non-zero orthonormal frame
components of $\sigma_{ab}$. On the other hand, the mean Hubble
volume expansion of the Lukash universe is determined by the
scalar
\begin{equation}
\Theta=\frac{1+2r}{t}\,.  \label{LTheta}
\end{equation}
The above ensures that the average scale factor obeys the simple
power law $a\propto t^{(1+2r)/3}$ with $0<r<1$. As expected, for
$r\rightarrow 1$ the scale factor evolution reduces to that of the
Milne universe (i.e.~$a\propto t$). Note that the effect of the
anisotropy is to reduce the expansion rate below that of the
isotropic case.

When measuring the average anisotropy of the expansion, it helps
to introduce the following dimensionless and expansion-normalised
shear parameter
\begin{equation}
\Sigma\equiv\frac{3\sigma^{2}}{\Theta^{2}}\,.  \label{Sigma}
\end{equation}
In the Lukash spacetime the scalars $\sigma^{2}$ and $\Theta$ are
given by (\ref{Lsigma2}) and (\ref{LTheta}) respectively. Using
these expressions we obtain
\begin{equation}
\Sigma=\frac{1-r}{1+2r}\,,  \label{LSigma}
\end{equation}
This result reflects the model's self similarity, which guarantees
that all the expansion-normalised dimensionless variables remain
constant in time. Given that $\Sigma>0$ and $0<r<1$, we
immediately deduce that $0<\Sigma<1,$ in accord with ${\cal R}<0$
in (\ref{LcR}). Thus, although the shear anisotropy is not
asymptotically stable (in the Lyapunov sense), it is stable in the
sense that any deviations from isotropy never
diverge~\cite{BS1}-\cite{BT}. Note that for $r\rightarrow1$ the
$\Sigma$-parameter approaches zero and the expansion becomes
isotropic. Maximum shear anisotropy (i.e.~$\Sigma\rightarrow1$),
on the other hand, corresponds to the $r=0$ limit.

Definition (\ref{Sigma}) also provides an alternative expression
for the Raychaudhuri equation of the Lukash solution. In
particular, combining (\ref{Ray1}) and (\ref{Sigma}) one arrives
at
\begin{equation}
\dot{\Theta}=-{\textstyle{\frac{1}{3}}}\Theta^{2}
\left(1+2\Sigma\right)\,,  \label{Ray2}
\end{equation}
while the power-law evolution of the average scale factor now
reads $a\propto t^{1/(1+2\Sigma)}$. Thus, in the absence of any
shear anisotropy we have $a\propto t$, as in the Milne universe.
For maximum shear anisotropy, on the other hand, we arrive at the
familiar scale-factor evolution of the Kasner vacuum solutions
(i.e.~$a\propto t^{1/3}$). Note that the deceleration parameter of
the Lukash model is $q=2\Sigma$, which means that $q=0$ when
$\Sigma=0$ and takes value $q=2$ of the Kasner points as
$\Sigma\rightarrow1$.

\subsection{Spatial curvature}
%%%%%%%%%%%%%%%%%%%%%%%%%%%%%%
In a vacuum, plane-wave Bianchi $VII{}_{h}$ spacetime, the trace
of the 3-Ricci tensor ${\cal R}_{\alpha\beta}$ associated with the
surfaces of constant time is
\begin{equation}
{\cal R}=-{\textstyle{\frac{1}{2}}}(n_{2}-n_{3})^{2}-
6a_{1}^{2}\,,  \label{LcR1}
\end{equation}
where $a_{1},\,n_{2}$ and $n_{3}$ are given in Eq.~(\ref{a1n2n3}).
According to the above ${\cal R}<0$ always, which guarantees the
hyperbolic geometry of the spatial sections. Note that we may use
(\ref{a1n2n3}) to recast Eq.~(\ref{LcR1}) as
\begin{equation}
{\cal R}=-\frac{k^{2}(1-f^{2})^{2}}{2f^{2}t^{2}}-
\frac{6r^{2}}{t^{2}}=-\frac{2r(1+2r)}{t^{2}}\,.  \label{LcR2}
\end{equation}
Then, combining results (\ref{Lsigma2}), (\ref{LTheta}) and
(\ref{LcR1}), one can show that that expression (\ref{LcR})
(i.e.~the Lukash analogue of the Friedmann equation) reduces to
the constraint (\ref{Lcons}a).

Spatial curvature anisotropies are described via the symmetric and
trace-free tensor ${\cal S}_{\alpha\beta}$. In the Lukash model,
the only non-zero components of ${\cal S}_{\alpha\beta}$ are
\begin{equation}
{\cal S}_{11}= -\frac{4r(1-r)}{3t^{2}}\,, \hspace{5mm} {\cal
S}_{22}= \frac{k^{2}(1-f^{2})(2+f^{2})}{3f^{2}t^{2}}\,,
\hspace{5mm} {\cal S}_{33}=
-\frac{k^{2}(1-f^{2})(1+2f^{2})}{3f^{2}t^{2}}  \label{LcS1}
\end{equation}
and
\begin{equation}
{\cal S}_{23}=-\frac{kr(1-f^{2})}{ft^{2}}\,.  \label{LcS2}
\end{equation}
According to (\ref{LcR2})-(\ref{LcS2}), the spatial curvature of
the model vanishes at the maximum shear limit, namely as
$r\rightarrow0$. At the other end, as $r$ approaches unity, only
the isotropic part of ${\cal R}_{\alpha\beta}$ survives. Recall
that $k^{2}(1-f^{2})=0$ as $r\rightarrow0$ or $1$
(see~Eq.~(\ref{Lcons}a)).

\subsection{Weyl curvature}
%%%%%%%%%%%%%%%%%%%%%%%%%%%
The only non-zero components of the Weyl curvature tensor
associated with a vacuum, plane-wave Bianchi type $VII{}_{h}$
spacetime are
\begin{equation}
E_{22}=-E_{33}=-H_{23}=\frac{k^{2}(1-f^{4})}{2f^{2}t^{2}}
\label{LWeyl1}
\end{equation}
and
\begin{equation}
H_{22}=-H_{33}=E_{23}=\frac{k(1-f^{2})(1-2r)}{2ft^{2}}\,.
\label{LWeyl2}
\end{equation}
This means that the Weyl curvature is minimised near the limits
$r=1$ and $r=0$, where $k^{2}(1-f^{2})=0$. Also, when $r=1/2$,
which corresponds to $\Sigma=1/4$ (see Eq.~(\ref{LSigma})), we
have $H_{22}=0=H_{33}=E_{23}$ and the Weyl tensor has only one
independent component.

These relations between the Weyl tensor components also guarantee
that the electric and magnetic Weyl tensors have equal magnitudes,
although they are orthogonal to each other. More specifically,
expressions (\ref{LWeyl1}) and (\ref{LWeyl2}) imply that
\begin{equation}
E^{2}=H^{2}=\frac{k^{2}(1-f^{2})^{2}}{4f^{2}t^{4}}
\left[\frac{k^{2}}{f^{2}}\left(1+f^{2}\right)^{2}
+(1+2r)^{2}\right]\,,  \label{E2=H2}
\end{equation}
and that
\begin{equation}
E_{\alpha\beta}H^{\alpha\beta}=0\,.  \label{EH=0}
\end{equation}

As with the expansion anisotropy, it helps to measure the
anisotropy of the Weyl field by means of the following
expansion-normalised, dimensionless scalars
\begin{equation}
{\cal W}_{+}\equiv\frac{W_{+}}{\Theta^{4}} \hspace{15mm} {\rm and}
\hspace{15mm} {\cal W}_{-}\equiv\frac{W_{-}}{\Theta^{4}}\,,
\label{WS}
\end{equation}
where $W_{\pm}=E^{2}\pm H^{2}$ by definition. The self-similarity
of the Lukash solution guarantees that ${\cal W}_{+}$ is time
independent, while the plane-wave nature of the model ensures that
${\cal W}_{-}=0$ (see constraint (\ref{E2=H2})).

\section{The perturbed Lukash solution}
%%%%%%%%%%%%%%%%%%%%%%%%%%%%%%%%%%%%%%%
The Lukash solution is the late-time attractor of the Bianchi
$VII{}_{h}$ spacetimes~\cite{BS1,BS2}, which are known to contain
the open FRW universe as a special subcase. In this respect,
studying the behaviour of the perturbed Lukash model could provide
useful clues to the final stages of ever-expanding FRW cosmologies
with $\mu+3p>0$. If $\mu+3p\leq0$ then the expansion will approach
the FRW (for power-law inflationary behaviour~\cite{no}) or de
Sitter universe in accord with the cosmic no-hair theorems.
Similar effects can arise from the effects of higher-order
curvature corrections to the Einstein-Hilbert lagrangian of
general relativity~\cite{cot}.

\subsection{Nonlinear equations}
%%%%%%%%%%%%%%%%%%%%%%%%%%%%%%%%
Consider a perturbed vacuum Bianchi $VII{}_{h}$ spacetime and
allow for a low-density, pressure-free matter component. The
nonlinear evolution of the latter is governed by the standard
energy-density conservation law
\begin{equation}
\dot{\mu}=-\Theta\mu\,.  \label{edc}
\end{equation}

When the matter component is in the form of dust, there is no
acceleration and the only additional kinematic contribution comes
from possible rotational disturbances. This means that the
cosmological velocity field, which is identified with the motion
of the matter, remains geodesic although it is allowed to rotate.
Rotation is monitored by the propagation equation of the vorticity
vector
\begin{equation}
\dot{\omega}_{a}=-{\textstyle{\frac{2}{3}}}\Theta\omega_{a}+
\sigma_{ab}\omega^{b}\,,  \label{dotomega}
\end{equation}
which also satisfies the constraint
\begin{equation}
{\rm D}^{a}\omega_{a}=0\,.  \label{divomega}
\end{equation}
According to (\ref{dotomega}), in addition to the expansion
effect, which always reduces vorticity, there is a contribution
due to the shear anisotropy. Note that for pressure-free matter
there are no sources of rotation and vorticity remains zero if it
was zero initially.

In the presence of a pressureless fluid and vorticity the
nonlinear Friedmann and Raychaudhuri equations respectively
give\footnote{In a low density perturbed model with $\Omega\ll1$,
the linear Friedmann and Raychaudhuri equations retain their
background functional form. The difference is that $\Sigma$ is
generally not constant, which means that the average scale factor
of the perturbed Lukash model no longer obeys the simple power-law
evolution given in section 3.2.}
\begin{equation}
{\cal R}=-{\textstyle{\frac{2}{3}}}\Theta^2(1-\Sigma-\Omega)-
2\omega^2  \label{nlFried}
\end{equation}
and
\begin{equation}
\dot{\Theta}=-{\textstyle{\frac{1}{3}}}\Theta^2
\left(1+2\Sigma+{\textstyle{\frac{1}{2}}}\Omega\right)+
2\omega^2\,,  \label{nlRay}
\end{equation}
where $\Omega=3\mu/\Theta^2$ is the density parameter. Similarly,
the introduction of matter and rotation modifies the rest of the
propagation formulae given in section 1.2 as follows
\begin{eqnarray}
\dot{E}_{ab}&=& -\Theta E_{ab}+ {\rm curl}H_{ab}+
3\sigma_{c\langle a}E_{b\rangle}{}^c-
{\textstyle{\frac{1}{2}}}\mu\sigma_{ab}-
\omega^c\epsilon_{cd\langle a}E_{b\rangle}{}^d\,, \label{nldotE}\\
\dot{H}_{ab}&=& -\Theta H_{ab}- {\rm curl}E_{ab}+
3\sigma_{c\langle a}H_{b\rangle}{}^d- \omega^c\epsilon_{cd\langle
a}H_{b\rangle}^d\,, \label{nldotH}\\ \dot{\sigma}_{ab}&=&
-{\textstyle{\frac{2}{3}}}\Theta\sigma_{ab}- E_{ab}-
\sigma_{c\langle a}\sigma^{c}{}_{b\rangle}- \omega_{\langle
a}\omega_{b\rangle}\,,  \label{nldotsigma}
\end{eqnarray}
while the associated constraints become
\begin{eqnarray}
H_{ab}&=& {\rm curl}\sigma_{ab}+ {\rm D}_{\langle
a}\omega_{b\rangle}\,, \label{nlcon1}\\ {\rm D}^b\sigma_{ab}&=&
{\textstyle{\frac{2}{3}}}{\rm D}_a\Theta+ {\rm curl}\omega_a\,,
\label{lcon2}\\ {\rm D}^bE_{ab}&=&
\epsilon_{abc}\sigma^b{}_dH^{cd}- 3H_{ab}\omega^b+
{\textstyle{\frac{1}{3}}}{\rm D}_a\mu\,,  \label{nlcon3}\\ {\rm
D}^bH_{ab}&=& -\epsilon_{abc}\sigma^b{}_dE^{cd}+ 3E_{ab}\omega^b+
\mu\omega_a\,.  \label{nlcon4}
\end{eqnarray}
Note that vorticity affects the expansion and the shear evolution
only at the nonlinear level, while it has a linear contribution in
Eqs.~(\ref{nldotE}), (\ref{nldotH}) and
(\ref{nlcon1})-(\ref{nlcon4}).

\subsection{Linear vortices}
%%%%%%%%%%%%%%%%%%%%%%%%%%%%
The presence of matter means that one can identify the
cosmological velocity field with that of the material component,
which in turn gives physical substance to the idea of rotation. We
measure the relative strength of rotational perturbations by means
of the expansion-normalised dimensionless scalar
\begin{equation}
\varpi=\frac{\omega}{\Theta}\,,  \label{vpi}
\end{equation}
with $\omega=(\omega_{a}\omega^{a})^{1/2}$. Taking the time
derivative of $\omega$ and using (\ref{Ray2}) and (\ref{dotomega})
we obtain the linear expression
\begin{equation}
\dot{\varpi}=-{\textstyle{\frac{1}{3}}}\tilde{\Theta}
\left(1-2\tilde{\Sigma}-3\tilde{\Sigma}_{ab}n^{a}n^{b}\right)
\varpi\,,  \label{lvptdot}
\end{equation}
where the tildas indicate background quantities. Here,
$\Sigma_{ab}=\sigma_{ab}/\Theta$ by definition and $n_{a}$ is the
unit vector along the rotation axis (i.e.~$\omega_{a}=\omega
n_{a}$). Written in an orthonormal frame the above reads
\begin{equation}
\partial_{{\rm t}}\varpi=-{\textstyle{\frac{1}{3}}}\tilde{\Theta}
\left(1-2\tilde{\Sigma}\right)\varpi+
\tilde{\Theta}\tilde{\Sigma}_{\alpha\beta}n^{\alpha}n^{\beta}
\varpi\,,  \label{lvpdot1}
\end{equation}
where $\tilde{\Sigma}_{\alpha \beta}$ and $n_{\alpha}$,
$n_{\beta}$ are the non-zero orthonormal frame components of
$\tilde{\Sigma}_{ab}$ and $n_{a}$, $n_{b}$ respectively. The first
term on the right-hand side of this equation describes the average
evolution of $\varpi $, while the second conveys the directional
effects. The former increases (or decreases) $\varpi$ depending on
whether $\tilde{\Sigma}$ is greater (or less) than 1/2.
\footnote{The value $\tilde{\Sigma}=1/2$, which corresponds to
$r=1/4$ (see expression (\ref{LSigma})), indicates a Lukash-type
spacetime with half the allowed amount of shear anisotropy. In the
case of rotational distortions, $\tilde{\Sigma}=1/2$ also
indicates the point where the background expansion rate drops
faster than the vorticity. Recall the decreasing effect of the
shear on the average expansion scalar (see Eqs.~(\ref{Ray1}),
(\ref{Ray2})). Therefore, at $\tilde{\Sigma}=1/2$ the scalar
$\varpi$ starts to increase on average (i.e.~excluding the
direction dependent effect of the last term in (\ref{lvpdot1})).}
Overall, linear vortices grow, relative to the average background
expansion, when the following condition holds
\begin{equation}
2\tilde{\Sigma}+
3\tilde{\Sigma}_{\alpha\beta}n^{\alpha}n^{\beta}>1\,.
\label{vcon}
\end{equation}
This condition implies that the growth of linear vortices also
depends on the relative orientation between the background shear
eigenframe and the rotation axis. Assuming that rotation takes
place along the ${\bf e}_{1}$ axis of the background orthonormal
frame, we may use expressions (\ref{Lsigma}a) and (\ref{LSigma})
to verify that condition (\ref{vcon}) holds as long as
$r<1/2$.\footnote{The condition $r<1/2$ on the metric parameter
translates into the constraint $\tilde{\Sigma}>1/4$ on the
background shear anisotropy. Accordingly, only three quarters of
the allowed Lukash backgrounds are potentially unstable to linear
rotational perturbations.} In this case linear vortices grow
relative to the background expansion as $\varpi\propto t^{1-2r}$.
This means that the growth rate of $\varpi$ takes the maximum
value $\varpi\propto t$ at the $r=0$ limit, namely for maximum
background shear. Alternatively, one may assume that the rotation
axis lies along ${\bf e}_{2}$ or ${\bf e}_{3}$. Then, a similar
calculation shows that linear vortices can never grow relative to
the average background expansion. At best, $\varpi$ remains
constant (when $r\rightarrow0$).

Following expressions (\ref{lvptdot}) and (\ref{lvpdot1}), the
average shear distortion always increases the residual amount of
rotation. Moreover, when condition (\ref{vcon}) is fulfilled, the
overall effect of the shear (including the direction dependent
component $\Sigma_{\alpha\beta}$) will also boost vorticity
perturbations. In other words, as far as rotation is concerned,
shear distortions can mimic the effects of matter pressure. Recall
that in the presence of pressure vorticity does not necessarily
decay with time (e.g.~see~\cite{E1,B2,BaTs}). Instead, for matter
with a stiff enough equation of state rotation will increase
despite the universal expansion. In our case, non-zero pressure
with $p=p(\mu)$ means that Eq.~(\ref{lvptdot}) takes the form
\begin{equation}
\dot{\varpi}=-{\textstyle{\frac{1}{3}}}\tilde{\Theta}
\left[1-2\tilde{\Sigma}-3\left(c_{{\rm s}}^{2}
+\tilde{\Sigma}_{ab}n^{a}n^{b}\right)\right]\varpi\,,
\label{lvptdot2}
\end{equation}
where $c_{{\rm s}}^{2}={\rm d}p/{\rm d}\mu$ is the square of the
adiabatic sound speed. This demonstrates clearly the analogy
between the shear and the pressure effects on rotation. For
example, when $\tilde{\Sigma}=1$ and the direction-dependent term
on the right-hand side of the above is negligible, the shear
effect on $\varpi$ is indistinguishable from that of a matter
component with $p/\mu=2/3$.

\subsection{Linear shear anisotropies}
%%%%%%%%%%%%%%%%%%%%%%%%%%%%%%%%%%%%%%
In a perturbed Bianchi~$VII{}_{h}$ model with low-density dust,
Raychaudhuri's formula (see Eq.~(\ref{nlRay})) ensures that the
linear expansion proceeds unaffected by the presence of matter or
by rotational distortions. Also, expression (\ref{nldotsigma})
guarantees that, to linear order, vortical perturbations do not
affect the evolution of the expansion-normalised shear parameter.
Thus, ignoring rotational and matter effects, we take the time
derivative of (\ref{Sigma}) and then use Eqs.~(\ref{Ray1}),
(\ref{LcSab}) and (\ref{dotsigma}) to arrive at
\begin{equation}
\dot{\Sigma}=-{\textstyle{\frac{4}{3}}}\Theta(1-\Sigma)\Sigma-
\frac{3}{\Theta^{2}}{\cal S}_{ab}\sigma^{ab}\,.  \label{ldotSigma}
\end{equation}
The first term in the right-hand side of this equation describes
the average linear evolution of $\Sigma$. Conversely, the last
term of Eq.~(\ref{ldotSigma}) describes directional effects and
depends on the relative orientations of the ${\cal S}_{ab}$ and
$\sigma_{ab}$ eigenframes. Relative to an orthonormal coordinate
system the above reads
\begin{equation}
\partial_{{\rm t}}\Sigma=
-{\textstyle{\frac{4}{3}}}\Theta(1-\Sigma)\Sigma-
\frac{3}{\Theta^{2}}{\cal S}_{\alpha\beta}\sigma^{\alpha\beta}\,,
\label{ldotSigma1}
\end{equation}
with ${\cal S}_{\alpha\beta}$ and $\sigma_{\alpha\beta}$
representing the non-zero orthonormal frame components of ${\cal
S}_{ab}$ and $\sigma_{ab}$ respectively.

Suppose that $\Sigma\rightarrow0$ to zero order. This state
corresponds to a background of minimum shear and 3-Ricci
anisotropy (i.e.~$r\rightarrow1$) with
$\tilde{\sigma}_{\alpha\beta}=0=\tilde{{\cal S}}_{\alpha\beta}$,
where tildas indicate the zero-order quantities. At this limit,
which corresponds to a perturbed Milne universe,
Eq.~(\ref{ldotSigma}) takes the linear form
\begin{equation}
\partial_{{\rm t}}\Sigma=
-{\textstyle{\frac{4}{3}}}\tilde{\Theta}\Sigma\,.
\label{ldotSigma2}
\end{equation}
As a result, $\Sigma\propto a^{-4}\propto t^{-4}$ to first order.
In other words, when $\tilde{\Sigma}\rightarrow0$ any linear
expansion anisotropies that may occur will quickly disperse.

When $\tilde{\Sigma}\rightarrow1$ we have $r\rightarrow0$ and the
background has maximum shear anisotropy and zero 3-Ricci
curvature. This means that $\tilde{{\cal S}}_{\alpha\beta}=0$ but
$\tilde{\sigma}_{\alpha\beta}\neq0$ (see expressions
(\ref{Lsigma}), (\ref{LcS1}) and (\ref{LcS2})). Here it helps to
introduce the auxiliary linear variable $S=1-\Sigma$ and rewrite
Eq.~(\ref{ldotSigma2}) as follows
\begin{equation}
\partial_{{\rm t}}S={\textstyle{\frac{4}{3}}}\tilde{\Theta}S+
\frac{3}{\tilde{\Theta}^{2}}\tilde{\sigma}_{\alpha\beta}{\cal
S}^{\alpha \beta }\,,  \label{ldotS}
\end{equation}
given that $\tilde{S}\rightarrow0$. Note that whenever $S$ grows
the $\Sigma$-parameter decreases and vice-versa. Also note that
although the background $S$ remains bounded within the open
interval $(0,\,1)$, this is not necessarily the case at the linear
level. Indeed, using the zero-order relations (\ref{Lsigma}) and
the trace-free nature of ${\cal S}_{\alpha\beta}$, we find that
the last term on the right-hand side of (\ref{ldotS}) equals
$3{\cal S}_{11}t$ in the $r\rightarrow0$ limit. This in turn
allows us to recast Eq.~(\ref{ldotS}) as
\begin{equation}
\partial_{{\rm t}}S=\frac{4}{3t}S+ 3{\cal S}_{11}t\,,
\label{ldotS1}
\end{equation}
where ${\cal S}_{11}$ is a component of the perturbed spatial
Ricci tensor. The above means that a decrease in $S$ and therefore
an increase in the linear shear anisotropy when
$\tilde{\Sigma}\rightarrow1$ is possible in principle. Suppose,
for example, that the perturbed ${\cal S}_{11}$ component retains
its background form, namely that ${\cal S}_{11}=-4r(1-r)/3t^{2}$
with $0<r\ll1$. Then, Eq.~(\ref{ldotS1}) solves to give
\begin{equation}
S=S_{0}\left(\frac{t}{t_{0}}\right)^{4/3}+
3r(1-r)\left[1-\left(\frac{t}{t_{0}}\right)^{4/3}\right]\,,
\label{lS}
\end{equation}
where $S_{0}=S(t_{0})$. Recalling that $S=1-\Sigma $ by
definition, this leads to the following expression for the
perturbed shear parameter:
\begin{equation}
\Sigma=1- 3r(1-r)+
\frac{3r^{2}(1-2r)}{1+2r}\left(\frac{t}{t_{0}}\right)^{4/3}\,,
\label{lSigma}
\end{equation}
assuming that $\Sigma_{0}=(1-r)/(1+2r)$. The latter means that the
initial relation between the perturbed $\Sigma$ and $r$ has the
background functional form. Given that $r\ll1$, expression
(\ref{lSigma}) implies that the perturbed $\Sigma$-parameter can
break through the $\Sigma=1$ barrier provided that
\begin{equation}
\left(\frac{t}{t_{0}}\right)^{4/3}>\frac{(1-r)(1+2r)}{r(1-2r)}\,.
\label{llSigma}
\end{equation}
For example, if $r\sim10^{-2}$, this will happen when
$t>10^{3/2}t_0$. Note, however, that the smaller the value of $r$
the longer it takes for the perturbed $\Sigma$ to cross through
unity.

It should be emphasised that $\Sigma>1$ to first order implies
that the linear 3-Ricci curvature of the perturbed Lukash model
becomes positive (see Eq.~(\ref{nlFried}) with
$\Omega=0=\omega_{a}$). This is possible for maximum background
shear (i.e.~as $r\rightarrow0$), because the zero-order spatial
Ricci tensor vanishes at that limit. So, in principle, the
perturbed spacetime can have spatial sections with slightly
positive curvature. Clearly, in this case the linear Lukash model
perturbs away from the family of the Bianchi~$VII{}_{h}$
cosmologies, a fact demonstrated by the unbounded shear parameter.
On the other hand, if we demand that the linear 3-curvature is
never positive, then $\Sigma$ will always remain bounded by unity.

\subsection{Linear Weyl anisotropy}
%%%%%%%%%%%%%%%%%%%%%%%%%%%%%%%%%%%
Consider the expansion-normalised dimensionless variable ${\cal
W}_{-}$ defined in (\ref{WS}a). This scalar vanishes in the Lukash
plane-wave background, which implies that a linear growth for
${\cal W}_{-}$ is a sign of instability at that perturbative
order. In addition, ${\cal W}_{-}$ directly determines the Weyl
curvature invariant $C_{abcd}C^{abcd}$. In this respect, the
linear evolution of ${\cal W}_{-}$ also monitors the gravitational
entropy of the perturbed Lukash universe and, to a certain extent,
that of low-density open FRW models.\footnote{A complete
description of the Weyl anisotropy of the perturbed vacuum Bianchi
$VII{}_{h}$ spacetime, also requires to study the linear evolution
of the expansion normalised scalar ${\cal W}_{+}$ (see definition
(\ref{WS}a)). Unlike ${\cal W}_{-}$, however, ${\cal W}_{+}$ has
nonzero background value. This complicates further the linear
study of ${\cal W}_{+}$ and allows only for relatively trivial
analytic solutions of the associated propagation equation. It is
conceivable that an improved version of the formalism presented
here will be able to address the full Weyl anisotropy of the
perturbed Lukash solution.}

For weakly rotating Lukash universes with a low-density dust
component (i.e.~when $\varpi,\,\Omega\ll1$), the time derivative
of definition (\ref{WS}b) leads to
\begin{eqnarray}
\dot{{\cal W}}_{-}&=&
-{\textstyle{\frac{2}{3}}}(1-4\Sigma)\Theta{\cal W}_{-}+
\frac{1}{\Theta^{4}}\left(E^{ab}{\rm curl}H_{ab}+H^{ab}{\rm
curl}E_{ab}\right) \nonumber\\
&{}&+\frac{3}{\Theta^{4}}\sigma_{ca}
\left(E^{c}{}_{b}E^{ab}-H^{c}{}_{b}H^{ab}\right)-
\frac{1}{\Theta^{4}}\omega^{c}\epsilon_{cda}
\left(E_{b}{}^{d}E^{ab}-H_{b}{}^{d}H^{ab}\right) \nonumber\\
&{}&-\frac{\mu}{2\Theta^{4}}\,\sigma_{ab}E^{ab}\,,
\label{ldotcW-}
\end{eqnarray}
on using Eqs.~(\ref{nlRay})-(\ref{nldotH}). The last two terms in
the right-hand side of the above describe the effects of vorticity
and matter respectively. Note that although matter does not
directly contribute to the Weyl field, the later is not entirely
arbitrary because of the contracted Bianchi identities. These are
in a sense the field equations for the Weyl curvature and, among
others, convey the matter effects on the propagation of the Weyl
components (e.g.~see~\cite{HE}). On introducing an orthonormal
frame, expression (\ref{ldotcW-}) reads
\begin{eqnarray}
\partial_{{\rm t}}{\cal W}_{-}&=&
-{\textstyle{\frac{2}{3}}}(1-4\Sigma)\Theta {\cal W}_{-}+
\frac{1}{\Theta^{4}}\epsilon_{\mu\nu\alpha}
\left(E^{\alpha\beta}\partial^{\mu}H^{\nu}{}_{\beta}
+H^{\alpha\beta}\partial^{\mu}E^{\nu}{}_{\beta}\right) \nonumber\\
&{}&-\frac{1}{\Theta^{4}}\epsilon_{\mu\nu\alpha}a^{\mu}
\left(E^{\alpha\beta}H^{\nu}{}_{\beta}
+H^{\alpha\beta}E^{\nu}{}_{\beta}\right)-
\frac{3}{\Theta^{4}}n_{\mu\alpha}
\left(E^{\alpha\beta}H^{\mu}{}_{\beta}
+H^{\alpha\beta}E^{\mu}{}_{\beta }\right) \nonumber \\
&{}&+\frac{1}{\Theta^{4}}n_{\mu}{}^{\mu}
E^{\alpha\beta}H_{\alpha\beta}+
\frac{3}{\Theta^{4}}\sigma_{\mu\alpha}
\left(E^{\mu}{}_{\beta}E^{\alpha \beta}
-H^{\mu}{}_{\beta}H^{\alpha\beta}\right) \nonumber \\
&{}&-\frac{1}{\Theta^{4}}\omega^{\mu }\epsilon_{\mu\nu\alpha}
\left(E_{\beta}{}^{\nu}E^{\alpha\beta}
-H_{\beta}{}^{\nu}H^{\alpha\beta}\right)-
\frac{\mu}{2\Theta^{4}}\,\sigma_{\alpha\beta}E^{\alpha\beta}\,.
\label{ldotcW-1}
\end{eqnarray}
Again, we notice that the first term on the right-hand side
describes the average evolution of ${\cal W}_{-}$, while the rest
describe direction-dependent effects. The former depends crucially
on the background shear anisotropy and it is reversed in sign at
$\tilde{\Sigma}=1/4$. This threshold corresponds to $r=1/2$ and
indicates the point where the background expansion rate starts
decreasing faster than the average $W_{-}$. Recall that at
$\tilde{\Sigma}=1/4$ the background Weyl field has only one
independent component (see section 3.4).

To proceed further we consider homogeneous perturbations in the
Weyl field, namely that
$\partial_{\mu}E_{\alpha\beta}=0=\partial_{\mu}H_{\alpha\beta}$.
This condition also monitors the large-scale behaviour of ${\cal
W}_{-}$ in the presence of inhomogeneities. In addition, we assume
that $\tilde{E}_{\alpha\beta}E_{\mu\nu}=
E_{\alpha\beta}\tilde{E}_{\mu\nu}$ and
$\tilde{H}_{\alpha\beta}H_{\mu\nu}=
H_{\alpha\beta}\tilde{H}_{\mu\nu}$ to linear order.\footnote{This
assumption, which allows us to obtain analytic solutions for the
linear evolution of ${\cal W}_{-}$, implies that the $E_{11}$ and
$H_{11}$ components of the perturbed model vanish. This
restriction already holds in the background.} Then, the combined
linear contribution of the second, third, fourth and fifth terms
in the right-hand side of (\ref{ldotcW-1}) is zero, while the
sixth term reduces to $-2(1-r){\cal W}_{-}/t$. Also, employing the
background relations between the Weyl tensor components given in
(\ref{LWeyl1}) and (\ref{LWeyl2}), one can immediately show that
the vorticity term in Eq.~(\ref{ldotcW-1}) vanishes to first
order. Finally, on using the zero-order expressions (\ref{Lcons}),
(\ref{Lsigma}), (\ref{LTheta}), (\ref{LWeyl1}) and given that
$\mu\propto t^{-(1+2r)}$ for dust (see Eq.~(\ref{edc})), we find
that
\begin{equation}
\frac{\mu}{2\tilde{\Theta}^{4}}\,
\tilde{\sigma}_{\alpha\beta}\tilde{E}^{\alpha\beta}=
\frac{\mu}{\tilde{\Theta}^{4}}\,
\tilde{\sigma}_{23}\tilde{E}_{23}\propto
\frac{r(1-r)(1-2r)}{(1+2r)^{4}}\,t^{-2r}\,,  \label{matter}
\end{equation}
to first order. Note that, according to this, the matter effect in
Eq.~(\ref{ldotcW-1}) vanishes when either $\tilde{\sigma}_{23}$ or
$\tilde{E}_{23}$ is zero. On these grounds the linear expression
(\ref{ldotcW-1}) reduces to
\begin{equation}
\partial_{{\rm t}}{\cal W}_{-}=-2r\,t^{-1}{\cal W}_{-}+
\frac{r(1-r)(1-2r)}{(1+2r)^{4}}\,{\cal C}\,t^{-2r}\,,
\label{ldotcW-2}
\end{equation}
where the parameter $r$ varies in the open interval $(0,\,1)$ and
${\cal C}$ is a constant. This result implies that the
expansion-normalised scalar ${\cal W}_{-}$ remains unchanged to
linear order as $r\rightarrow0$, that is for maximum background
shear. The reason is that in the $r=0$ limit the Weyl anisotropy
of the Lukash solution disappears (see Eqs.~(\ref{LWeyl1}),
(\ref{LWeyl2})). The matter effect also vanishes near the minimum
background shear limit (i.e.~as $r\rightarrow1$) and at the
$r=1/2$ threshold. The former of these two results is not
surprising, since the Lukash solution decays to the Milne universe
at the $r=1$ limit. When $r=1/2$, however, the matter effect is
zero because $\tilde{E}_{23}=0$ at that point (see
Eq.~(\ref{LWeyl2})). Recall that at the $r=1/2$ threshold the Weyl
tensor has only one independent component.

When $r\rightarrow0$ we find that $\partial_{{\rm t}}{\cal
W}_{-}=0$, which implies that any deviations in ${\cal W}_{-}$
that may occur will remain constant. On the other hand, as
$r\rightarrow1$ or at the $r=1/2$ threshold the
expansion-normalised Weyl parameter decays as
\begin{equation}
{\cal W}_{-}\propto t^{-2} \hspace{10mm} {\rm and} \hspace{10mm}
{\cal W}_{-}\propto t^{-1}\,,  \label{lcW-1}
\end{equation}
respectively. In general, ignoring the effect of matter leads to
${\cal W}_{-}\propto t^{-2r}$ and therefore ensures that ${\cal
W}_{-}$ dies away at a rate inversely proportional to the
background shear anisotropy. This in turn implies that
$W_{-}\propto t^{-2(2+r)}$, since $\tilde{\Theta}\propto t^{-1}$.
The linear decay of the ${\cal W}_{-}$ parameter is in agreement
with the stability of the vacuum, plane-wave equilibrium points
that consist the late time asymptotes of the Bianchi $VII_h$
spacetimes (e.g.~see~\cite{HW1,BS2,HvdHC}).

The situation changes in the presence of matter. The latter, even
when it is in the form of a low-density dust component, introduces
new degrees of freedom into the system and the decrease of ${\cal
W}_{-}$ is not always guaranteed. Indeed, the general solution of
Eq.~(\ref{ldotcW-2}) reads
\begin{equation}
{\cal W}_{-}= {\cal C}_{1}t^{1-2r}+ {\cal C}_{2}t^{-2r}\,,
\label{lcW-}
\end{equation}
where ${\cal C}_{1}$, ${\cal C}_{2}$ are constants and ${\cal
C}_{1}={\cal C}r(1-r)(1-2r)/(1+2r)^{4}$. According to the above,
the expansion-normalised scalar ${\cal W}_{-}$ will start
increasing as ${\cal W}_{-}\propto t^{1-2r}$ when $0<r<1/2$. The
latter corresponds to $\tilde{\Sigma}>1/4$, which means that three
quarters of the allowed Lukash models, those with the largest
background shear anisotropy, are unstable against linear Weyl
curvature distortions. Incidentally, the same family was also
found vulnerable to linear rotational distortions (see section
5.2). Finally, we note that the aforementioned matter effects are
sensitive to the precise evolution of the background model, namely
to the properties of the Lukash vacuum solution. Given that, one
should be careful before extrapolating these results to Bianchi
$VII_h$ models with dust.

\section{Discussion}
%%%%%%%%%%%%%%%%%%%%
Bianchi models, particularly those that contain the FRW
cosmologies as special subcases, are essential for our
understanding of the large scale anisotropy of the universe. In
the family of Bianchi spacetimes, those of type $VII{}_{h}$ are
the most general homogeneous models containing the spatially open
FRW universe. In the absence of matter these spacetimes reduce to
the plane-wave solution found by Doroshkevich and
Lukash~\cite{DLN,L2}. These vacuum models also act as the future
attractors for the non-tilted, perfect fluid Bianchi $VII{}_{h}$
cosmologies~\cite{HW2}. For this reason the empty $VII{}_{h}$
Lukash model has been used to study the late-time evolution of
perturbed open FRW universes with a conventional matter content.

Here we have engaged a mixture of covariant and orthonormal frame
methods to study analytically the linear response of the Lukash
solution to a variety of perturbations. Our results show that the
amount of the background shear anisotropy is crucial for the
stability of the vacuum model. More specifically, by looking into
rotational or shear perturbations, we found that the linear
instability of these distortions is more likely in models with
higher background shear anisotropy. When dealing with vorticity
perturbations, in particular, the background shear can force
linear rotational perturbations to grow, thus mimicking the
effects of a fluid with a stiff equation of state. Also, when the
unperturbed model has the maximum allowed kinematical anisotropy,
linear shear distortions are no longer necessarily bounded. In the
latter case the Lukash universe perturbs away from the family of
the Bianchi~$VII{}_h$ spacetimes.

Our linear analysis also considered the effects of a non-zero
matter component. When the latter was in the form of low-density
dust, we found that matter had no effect on either the rotation or
the kinematical anisotropy of the perturbed Lukash solution.
However, the presence of matter (even a non-relativistic
pressureless fluid) plays a key role in the evolution of the Weyl
anisotropy of the perturbed spacetime. In particular, our study
showed that the introduction of a pressure-free component at the
linear level is the catalyst that can force the vacuum $VII{}_{h}$
model to diverge from its original plane-wave nature.

In the present paper the study of the matter effects has been
confined to the case of a pressure-free component. It is
relatively straightforward to extend this formalism to include the
effects of pressure. Generally speaking, nonzero pressure means
that the matter term in Eq.~(\ref{ldotcW-2}) decays faster than in
the case of dust. This in turn should make the matter effects on
${\cal W}_{-}$ less pronounced. However, this rather intuitive
picture will be probably complicated by the presence of pressure
gradients, which for dust are identically zero. One could also
consider the potential implications of large-scale magnetic fields
or of a non-conventional, `dark' matter component. The study of
magnetic fields in Bianchi $VII$, for example, has lead to limits
on the strength of a possible large-scale homogeneous field more
stringent than those obtained from standard nucleosynthesis
constraints~\cite{mag}. This happens because anisotropic stresses
play an important role in the evolution of simple anisotropic
universes of Bianchi type I. The anisotropic trace-free stress
mimics the part played by the anisotropic curvature in type $VII$
models and, in combination with a perfect fluid, slows the decay
of the shear anisotropy in a subtle
way~\cite{aniso}-\cite{aniso2}. The latter leads to more severe
observational consequences for the CMB. A further consideration
for future work is the role of dark energy in the universe. For a
perfect fluid with $\mu+3p<0$, in violation of the strong-energy
condition, the Lukash metric is unstable and approaches the flat
Friedmann universe as $t\rightarrow\infty$, following with the
course of power-law inflation. If $p=-\mu$, then the dynamics
approach the de Sitter universe with exponential rapidity within
the event horizon of any geodesically moving
observer~\cite{wald}-\cite{coley}. This case is less interesting
from a mathematical point of view because all distortions are
rapidly inflated away. However, it is likely to be of considerable
astronomical interest because of the growing observational
evidence that inflation has played some role in the very early
evolution of the universe and that dark energy, with $\mu+3p<0$,
is dominating the dynamics of the universe again today. Finally,
we note that if the universe is negatively curved with compact
topology there are severe constraints on the possibility of any
homogeneous anisotropy existing in the expansion at all: all
Bianchi $VII_{h}$ universes have to be
isotropic~\cite{ash}-\cite{bk2}.

\section*{Acknowledgements}
%%%%%%%%%%%%%%%%%%%%%%%%%%%
The authors would like to thank Pantelis Apostolopoulos, Alan
Coley, Sigbj\o rn Hervik, Henk van Elst and John Wainwright for
helpful discussions and comments.

\end{document}